\documentclass{aa}
\usepackage{graphicx}
\usepackage{longtable}
\usepackage{txfonts}

\newcommand{\hess}{H.E.S.S.$\,$}

\begin{document}

%%\title{Very high energy $\gamma$-ray observations of the starburst
  %%galaxy NGC\,253 with \hess}
\title{A search for very high energy $\gamma$-ray emission from the starburst galaxy NGC\,253 with \hess}

\titlerunning{A search for VHE $\gamma$-ray emission from NGC\,253 with \hess}

\author{F. Aharonian\inst{1}
 \and A.G.~Akhperjanian \inst{2}
 %\and K.-M.~Aye \inst{7}
 \and A.R.~Bazer-Bachi \inst{3}
 \and M.~Beilicke \inst{4}
 \and W.~Benbow \inst{1}
 \and D.~Berge \inst{1}
 %\and P.~Berghaus \inst{9} \thanks{Universit\'e Libre de 
 %Bruxelles, Facult\'e des Sciences, Campus de la Plaine, CP230, Boulevard
 %du Triomphe, 1050 Bruxelles, Belgium}
 \and K.~Bernl\"ohr \inst{1,5}
 \and C.~Boisson \inst{6}
 \and O.~Bolz \inst{1}
 \and V.~Borrel \inst{3}
 \and I.~Braun \inst{1}
 \and F.~Breitling \inst{5}
 \and A.M.~Brown \inst{7}
 %\and J.~Bussons Gordo \inst{12}
 \and P.M.~Chadwick \inst{7}
 \and L.-M.~Chounet \inst{8}
 \and R.~Cornils \inst{4}
 \and L.~Costamante \inst{1,20}
 \and B.~Degrange \inst{8}
 \and H.J.~Dickinson \inst{7}
 \and A.~Djannati-Ata\"i \inst{9}
 \and L.O'C.~Drury \inst{10}
 \and G.~Dubus \inst{8}
 \and D.~Emmanoulopoulos \inst{11}
 \and P.~Espigat \inst{9}
 \and F.~Feinstein \inst{12}
 %\and P.~Fleury \inst{8}
 \and G.~Fontaine \inst{8}
 \and Y.~Fuchs \inst{13}
 \and S.~Funk \inst{1}
 \and Y.A.~Gallant \inst{12}
 \and B.~Giebels \inst{8}
 \and S.~Gillessen \inst{1}
 \and J.F.~Glicenstein \inst{14}
 \and P.~Goret \inst{14}
 \and C.~Hadjichristidis \inst{7}
 \and M.~Hauser \inst{11}
 \and G.~Heinzelmann \inst{4}
 \and G.~Henri \inst{13}
 \and G.~Hermann \inst{1}
 \and J.A.~Hinton \inst{1}
 \and W.~Hofmann \inst{1}
 \and M.~Holleran \inst{15}
 \and D.~Horns \inst{1}
 \and A.~Jacholkowska \inst{12}
 \and O.C.~de~Jager \inst{15}
 \and B.~Kh\'elifi \inst{1}
 \and Nu.~Komin \inst{5}
 \and A.~Konopelko \inst{1,5}
 \and I.J.~Latham \inst{7}
 \and R.~Le Gallou \inst{7}
 \and A.~Lemi\`ere \inst{9}
 \and M.~Lemoine-Goumard \inst{8}
 \and N.~Leroy \inst{8}
 \and T.~Lohse \inst{5}
 \and J.M.~Martin \inst{6}
 \and O.~Martineau-Huynh \inst{16}
 \and A.~Marcowith \inst{3}
 \and C.~Masterson \inst{1,20}
 \and T.J.L.~McComb \inst{7}
 \and M.~de~Naurois \inst{16}
 \and S.J.~Nolan \inst{7}
 \and A.~Noutsos \inst{7}
 \and K.J.~Orford \inst{7}
 \and J.L.~Osborne \inst{7}
 \and M.~Ouchrif \inst{16,20}
 \and M.~Panter \inst{1}
 \and G.~Pelletier \inst{13}
 \and S.~Pita \inst{9}
 \and G.~P\"uhlhofer \inst{1,11}
 \and M.~Punch \inst{9}
 \and B.C.~Raubenheimer \inst{15}
 \and M.~Raue \inst{4}
 \and J.~Raux \inst{16}
 \and S.M.~Rayner \inst{7}
% \and I.~Redondo \inst{8,20}\thanks{now at Department of Physics and
%Astronomy, Univ. of Sheffield, The Hicks Building,
%Hounsfield Road, Sheffield S3 7RH, U.K.}
 \and A.~Reimer \inst{17}
 \and O.~Reimer \inst{17}
 \and J.~Ripken \inst{4}
 \and L.~Rob \inst{18}
 \and L.~Rolland \inst{16}
 \and G.~Rowell \inst{1}
 \and V.~Sahakian \inst{2}
 \and L.~Saug\'e \inst{13}
 \and S.~Schlenker \inst{5}
 \and R.~Schlickeiser \inst{17}
 \and C.~Schuster \inst{17}
 \and U.~Schwanke \inst{5}
 \and M.~Siewert \inst{17}
 \and H.~Sol \inst{6}
 \and D.~Spangler \inst{7}
 \and R.~Steenkamp \inst{19}
 \and C.~Stegmann \inst{5}
 \and J.-P.~Tavernet \inst{16}
 \and R.~Terrier \inst{9}
 \and C.G.~Th\'eoret \inst{9}
 \and M.~Tluczykont \inst{8,20}
 \and G.~Vasileiadis \inst{12}
 \and C.~Venter \inst{15}
 \and P.~Vincent \inst{16}
 \and H.J.~V\"olk \inst{1}
 \and S.J.~Wagner \inst{11}}

\offprints{J.A. Hinton, \email{Jim.Hinton@mpi-hd.mpg.de}}

\institute{
Max-Planck-Institut f\"ur Kernphysik, Heidelberg, Germany
\and
 Yerevan Physics Institute, Armenia
\and
Centre d'Etude Spatiale des Rayonnements, CNRS/UPS, Toulouse, France
\and
Universit\"at Hamburg, Institut f\"ur Experimentalphysik, Germany
\and
Institut f\"ur Physik, Humboldt-Universit\"at zu Berlin, Germany
\and
LUTH, UMR 8102 du CNRS, Observatoire de Paris, Section de Meudon, France
\and
University of Durham, Department of Physics, U.K.
\and
Laboratoire Leprince-Ringuet, IN2P3/CNRS, Ecole Polytechnique, Palaiseau, France
\and
%Physique Corpusculaire et Cosmologie, IN2P3/CNRS, Coll{\`e}ge de France, 11 Place
%Marcelin Berthelot, F-75231 Paris Cedex 5, France
APC, Paris, France 
(UMR 7164, CNRS, Universit\'e Paris VII, CEA, Observatoire de Paris)
\and
Dublin Institute for Advanced Studies, Ireland
\and
Landessternwarte, K\"onigstuhl, Heidelberg, Germany
\and
Laboratoire de Physique Th\'eorique et Astroparticules, IN2P3/CNRS,
Universit\'e Montpellier II, France
\and
Laboratoire d'Astrophysique de Grenoble, INSU/CNRS, Universit\'e
Joseph Fourier, France 
\and
DAPNIA/DSM/CEA, CE Saclay, Gif-sur-Yvette, Cedex, France
\and
Unit for Space Physics, North-West University, Potchefstroom, South Africa
\and
Laboratoire de Physique Nucl\'eaire et de Hautes Energies, IN2P3/CNRS, Universit\'es
Paris VI \& VII, Paris, France
\and
Institut f\"ur Theoretische Physik, Lehrstuhl IV: Weltraum und
Astrophysik, Ruhr-Universit\"at Bochum, Germany
\and
Institute of Particle and Nuclear Physics, Charles University,
Prague, Czech Republic
\and
University of Namibia, Windhoek, Namibia
\and
European Associated Laboratory for Gamma-Ray Astronomy, jointly
supported by CNRS and MPG}

\date{Received ? / Accepted ?}

\abstract{

We present the result of 28 hours of observations of the nearby 
starburst galaxy NGC~253
with the \hess detector in 2003. We find no evidence for very high energy 
$\gamma$-ray emission from this object. Gamma-ray emission above 400~GeV 
from NGC~253 had been reported by the CANGAROO collaboration in 2002. 
From the \hess data we derive upper limits on the flux above 300~GeV of $1.9\,\times\,10^{-12}$
photons~cm$^{-2}$ s$^{-1}$ for a point-like source and $6.3\,\times\,10^{-12}$
photons~cm$^{-2}$ s$^{-1}$ for a source of radius 0.5$^{\circ}$ as reported by
CANGAROO, both at a confidence level of 99\%. These upper limits are
inconsistent with the spectrum reported by CANGAROO. 
The expected very high energy $\gamma$-ray emission from this object is
discussed in the framework of a galactic wind propagation model.

;

}

\maketitle

%_________________________________________________________________
\section{Introduction}
\label{intro}

%%% Add similarity of M 82 and NGC 253 - Hamburg
Starburst galaxies, such as the closest objects of this type, NGC~253
in the Southern and M~82 in the Northern Hemisphere, are known to
exhibit enhanced and strongly localised supernova explosion rates in
so called 'starburst regions'.  In such regions gas and photon
densities are very high compared to the situation in normal star
forming galaxies like the Milky Way.  Given the paradigm of cosmic-ray
(CR) acceleration in supernova remnants (SNR), substantially greater
CR energy densities $E_{\mathrm{c}}\geq 100\times
E_{\mathrm{c}}^{\mathrm{gal}}$ (where
$E_{\mathrm{c}}^{\mathrm{gal}}\sim 1\,\mathrm{eV}\,\mathrm{cm}^{-3}$
is the value in our galaxy) are expected in such regions, with
corresponding fluxes for $\pi^{0}$-decay $\gamma$-rays alone in the
range of $10^{-9} \leq F_{\gamma}(>300\,\mathrm{MeV}) \leq 10^{-8}$
photons cm$^{-2}$ s$^{-1}$ for an object such as M~82
(V\"olk~et~al.~\cite{Voelk1}).  Later estimates for
M\,82~(Aky\"uz~et~al.~\cite{Aky}; V\"olk~et~al.~\cite{Voelk2}) are
approximately encompassed by this range, which translates into a
$\gamma$-ray flux of $1.5\,\times\,10^{-13} \leq F_{\gamma}( > 1
\mathrm{TeV})\times (E_{\gamma}/1 \mathrm{TeV})^{1.1} \leq
1.5\,\times\,10^{-12}$~photons cm$^{-2}$ s$^{-1}$~at energies
$E_{\gamma} \gg 300$~MeV assuming a differential proton energy
spectrum $\propto E_{p}^{-2.1}$.

EGRET observations of NGC~253 (Sreekumar~et~al.~\cite{Sreekumar}), the
nearest spiral galaxy outside the local group (at distance $d\approx
2.6$~Mpc), resulted in an upper limit on the integral flux above 100
MeV of $1.0 \times 10^{-7}$ photons cm$^{-2}$ s$^{-1}$. A reanalysis
of these data by Paglione~et~al.~(\cite{Paglione}) yielded a
consistent limit of $8 \times 10^{-8}$ photons cm$^{-2}$
s$^{-1}$. More complete EGRET data led Blom~et~al.~(\cite{Blom}) to
revise this upper limit down to $3.4 \times 10^{-8}$ photons cm$^{-2}$
s$^{-1}$; for M\,82 they derive an upper limit of $4.4 \times 10^{-8}$
photons cm$^{-2}$ s$^{-1}$.  Similar upper limits for both NGC~253 and
M~82 were obtained in a recent reanalysis of the EGRET data by
Cillis~et~al.~(\cite{Cillis}).  NGC~253 was detected in the hard X-ray
band using the OSSE instrument (Bhattacharya~et~al.~\cite{OSSE}). From
the 4.4$\sigma$ significance detection a flux of $3 \times 10^{-11}$
erg cm$^{-2}$ s$^{-1}$ (50-165~keV) was derived.

At TeV energies, observations of both NGC~253 and M~82 were made by
the HEGRA collaboration, but neither object was
detected~(G\"otting~\cite{HEGRA}).  For NGC~253 a flux upper limit of
$F_{\gamma}(>5.2\,\mathrm{TeV}) < 1.3 \times 10^{-13}$ photons
cm$^{-2}$ s$^{-1}$ at 99\% confidence was derived from 32.5 hours of
live time. For M~82 the corresponding limit was
$F_{\gamma}(>2.1\,\mathrm{TeV}) < 2.7 \times 10^{-13}$ photons
cm$^{-2}$ s$^{-1}$ from 43.9 hours of live time.  The report of very
high energy (VHE, used here as 0.1~TeV -- 10~TeV) $\gamma$-ray
emission from NGC\,253 by the CANGAROO collaboration
(Itoh~et~al.~\cite{CANGAROODetection}) marked the first claim of
$\gamma$-ray emission from a starburst galaxy. The differential energy
spectrum (Itoh~et~al.~\cite{CANGAROOEvidence}) can be described by:
$dF/dE =
(2.85\,\pm\,0.71)\,\times\,10^{-12}\,\times\,(E/1\,\mathrm{TeV})^{-3.85\,\pm\,0.46}$
cm$^{-2}$ s$^{-1}$ TeV$^{-1}$, with a corresponding flux above 400 GeV
of $\approx\,1.4\,\times\,10^{-11}$ cm$^{-2}$ s$^{-1}$, or
$\approx\,15$\% of the flux of the Crab Nebula above the same
threshold.  The emission reported by CANGAROO is extended with an rms
angular size estimated as 0.3--0.6$^{\circ}$, somewhat larger than the
optical size of the galaxy (28$'$ $\times$ 7$'$).

Further evidence for particle acceleration in NGC~253, at lower photon
energies, exists mainly in the form of radio data. Radio synchrotron
observations show an extended, quasi-spherical halo, rising to at
least 9 kpc above the disk (Carilli~et~al.~\cite{Carilli}), with a
morphology that suggests a gas outflow from the disk, carrying with it
magnetic fields and radiating electrons. Such an outflow should be
characteristic for starburst galaxies.  It has been studied in detail
in thermal soft X-rays (e.g.  Dahlem~et~al.~\cite{Dahlem},
Strickland~et~al.~\cite{Strick1}, \cite{Strick2}, \cite{Strick3})~and
interpreted in terms of ''superwinds``. The observed X-ray emission
shows a good correlation with optical $\mathrm{H}_{\alpha}$~line and
radio continuum observations over the same region.  From general
considerations of particle acceleration, it seems likely that
accelerated nuclear particles are also present and contribute to
driving the wind.  The CANGAROO detection has been interpreted as
inverse Compton emission, implying the existence of an extended halo
of multi-TeV~CR electrons around NGC~253
(Itoh~et~al.~\cite{CANGAROOHalo}). A cold dark matter (CDM)
annihilation interpretation was also discussed
(Enomoto~et~al.~\cite{Enomoto}).

The detection of NGC~253 by CANGAROO, along with the theoretical
expectation of very high energy $\gamma$-rays from starburst galaxies,
motivated the observation of this object with the High Energy
Stereoscopic System (\hess).  \hess is an array of four 13~m diameter
telescopes, each equipped with 960 pixel (5$^{\circ}$ field-of-view)
cameras (Hinton~\cite{HESSProject}).  Commissioning of the array in
Namibia (at 1800~m above sea level) began in 2002 and the array was
completed in December 2003. The optical system of the \hess telescopes
is described in detail by Bernl\"ohr~et~al.~(\cite{HESSOptics}) and
Cornils~et~al.~(\cite{HESSOptics2}), the trigger system by
Funk~et~al.~(\cite{HESSTrigger}) and the camera electronics by
Vincent~et~al.~(\cite{HESSCamera}). Observations of NGC~253, made
during 2003 with \hess, are presented here.

\section{\hess Observations}

Observations of NGC~253 were made during the construction
of the \hess array. In August and September 2003, 30~hours 
of data were taken with two telescopes in operation. In October
2003 a further 8~hours of 3-telescope data were taken. After
run quality selection (primarily to discard data compromised
by unstable weather conditions) and dead time correction, datasets
of 24.6~hours (August/September) and 3.4~hours (October) remain.
The mean zenith angle of observations was 14$^{\circ}$.
Only events where at least two telescopes provided shower images were
used in this analysis to enable stereoscopic reconstruction.
Data were taken in {\it wobble mode}, with the 
target source offset by 0.5$^{\circ}$ from the tracking position
of the telescopes. The energy threshold for this dataset is 190~GeV.
Two independent analysis methods have been applied to search for $\gamma$-ray
emission in this dataset.

\subsection{Analysis Method 1}

The standard calibration and analysis used for H.E.S.S. is described
in detail in (Aharonian~et~al.~\cite{HESS2155};
Aharonian~et~al.~\cite{HESSCalib}) and is referred to here as \emph{Analysis~1}.
Following calibration, \emph{tail-cuts} image cleaning and image moment
analysis (Hillas parameter determination) were used to characterise
each telescope image. Shower direction reconstruction was made based on 
the intersection of the major axes of the individual telescope images.
The angular resolution (defined by the radius containing 68\% of the
reconstructed events from a point-source) is 0.14$^{\circ}$
for the two telescope data set and 0.12$^{\circ}$ for three telescopes.
The primary $\gamma$-ray energy is estimated (with a resolution of
$\approx$\,15\%) using the total image size in each telescope and 
the reconstructed shower core position.
The mean reduced scaled width and length of shower images are used 
to select $\gamma$-ray candidate events. 

\begin{figure}
\centering
\includegraphics[width=8.7cm]{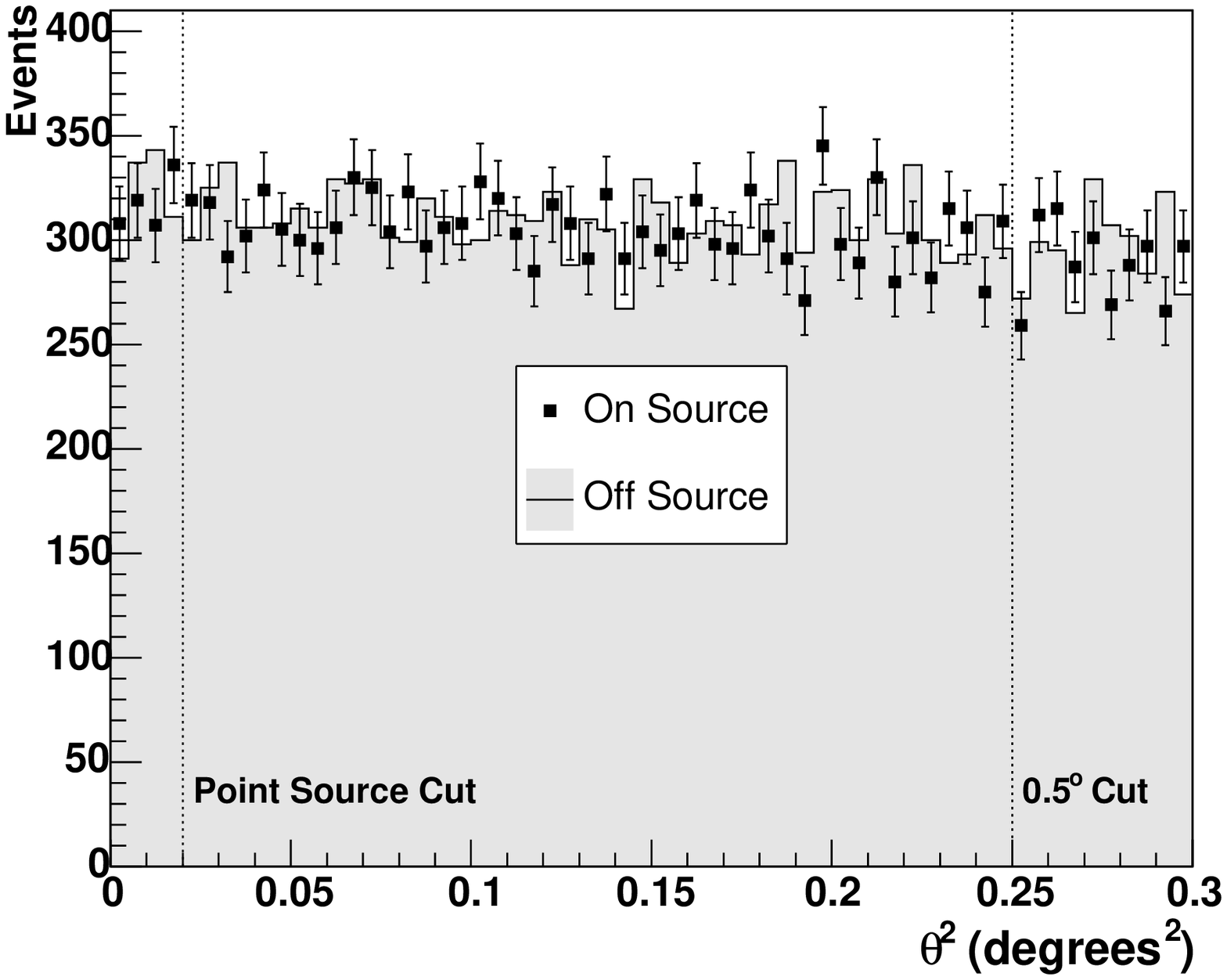}\vspace{-2mm}\hspace{1mm}
\includegraphics[width=8.7cm]{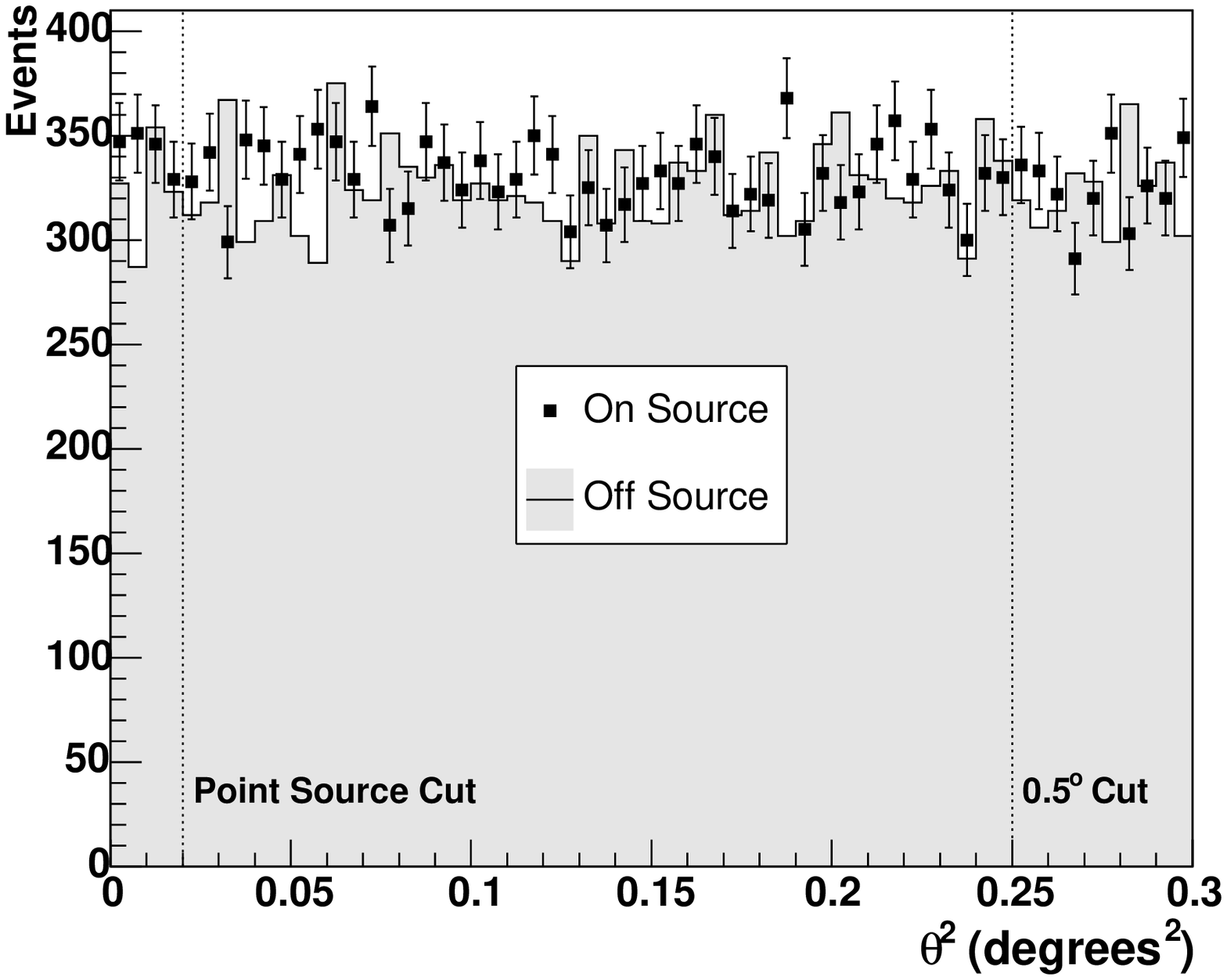}
\caption{
  Top: Angular distribution of $\gamma$-like images relative to the centre of
  NGC~253 (``On'') and relative to the background control region
  (``Off'') for Analysis~1. Events are plotted versus the squared angular distance to give equal solid
  angle in each bin. Background curves (histograms) are determined relative to 
  to points $1^{\circ}$ away from the source position. 
  Bottom: as for upper figure 
  but using Analysis~2 (described in section~\ref{sec:analysis2}).
} 
\label{fig:thetasq}
\end{figure}

Given that the observations were taken in wobble-mode, the position
0.5$^{\circ}$ from the tracking position on the opposite side of the
field of view (FOV) can be used for a simultaneous background
measurement. Figure~\ref{fig:thetasq} shows the distribution of
squared angular distance ($\theta^{2}$) of candidate $\gamma$-rays
from the centre of NGC~253 for the combined 2- and 3-telescope data
sets. The distributions for the on-source and background measurements
are consistent; no evidence for a $\gamma$-ray signal is seen. The
vertical dashed lines in this figure indicate the positions of the
standard \hess\ $\gamma$-ray selection cut for point-like sources
($\theta < 0.14^{\circ}$) and a much wider cut (at 0.5$^{\circ}$) at
the approximate extent reported by the CANGAROO collaboration.

\begin{figure}
\centering
\includegraphics[height=7.8cm]{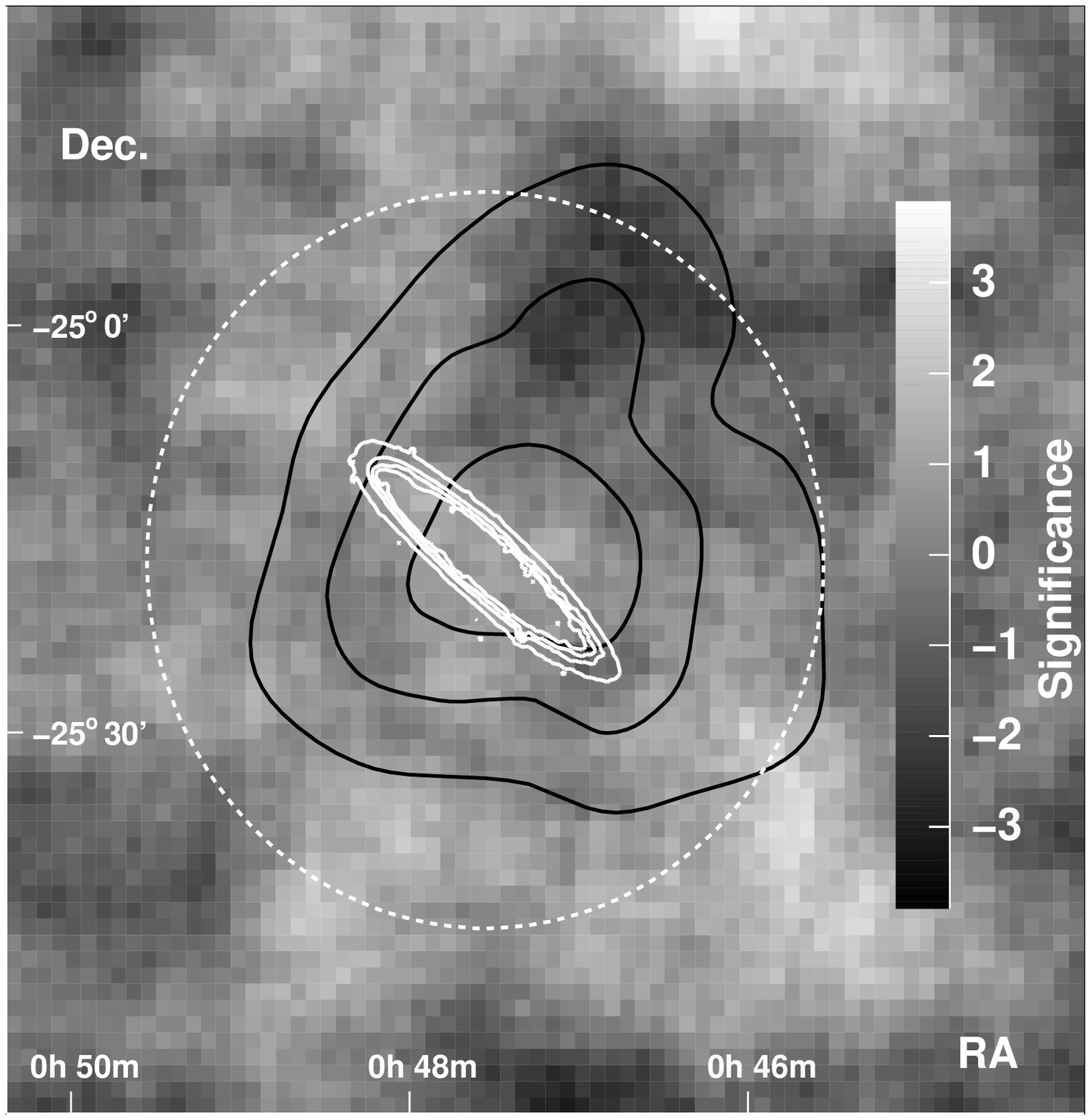}\vspace{2mm}\hspace{1mm}
\includegraphics[height=7.8cm]{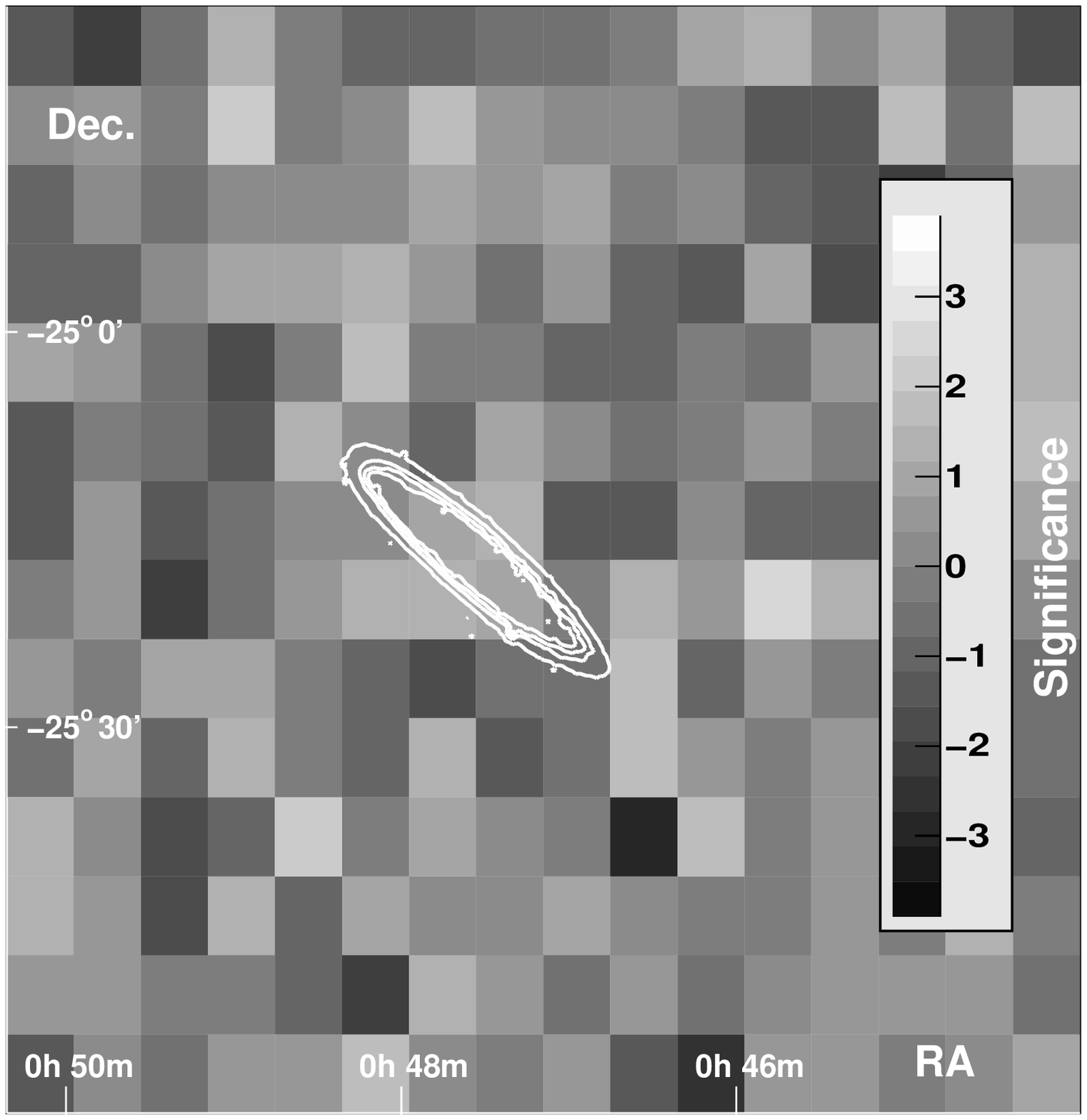}
\caption{
  Top: $\gamma$-ray point source significance map (grey-scale) derived from two
  telescope \hess\ data taken in August and September 2004
  and using Analysis~1. 
  The white contours show the optical emission from the galaxy (in a linear scale).
  These data are taken from the STScI Digitized Sky Survey~\cite{Optical}.
  Black contours are confidence levels (at 40\%, 65\% and 80\%) 
  for TeV $\gamma$-ray emission reported by
  Itoh~et~al.~(\cite{CANGAROODetection}). 
  The dashed white line shows 
  the angular cut used to derive the extended source flux limits
  presented here. In this approach bins are correlated within the
  radius of the $\theta$ cut (0.14$^{\circ}$). 
  Bottom: Significance map derived from the same dataset using Analysis~2.
  In contrast to the Top panel the statistical significance in each bin
  is independent. The bin size is matched to the angular resolution of
  the instrument for these data.
}
\label{fig:skymap}
\end{figure}

To enable 2-dimensional mapping of the FOV a different background
model is applied. The upper panel of figure~\ref{fig:skymap} shows the 
significance of a point-like excess in the FOV of NGC~253 using a 
ring (of 0.5$^{\circ}$ radius) around each position to provide a 
background estimate (a correction for the radial acceptance function 
of the instrument is made in this case). The distribution of
significance within the FOV is consistent with random 
fluctuations of the background. 

\subsection{Analysis Method 2}
\label{sec:analysis2}

An independent analysis method for the H.E.S.S. data
(Lemoine-Goumard~et~al.~\cite{HESS3DModel}, \emph{Analysis 2}) has
been used to confirm this result. This analysis employs not only
different reconstruction and background reduction methods but also an
independent detector calibration
(Aharonian~et~al.~\cite{HESSCalib}). The reconstruction method is
based on a simple 3D-modeling of an electromagnetic shower assuming
rotational symmetry. For each event eight parameters are
reconstructed: the direction of the shower in the reference frame of
the telescope, the core position on the ground, the altitude of shower
maximum, the total number of emitted Cherenkov photons, and the
longitudinal and transverse standard deviations of the Gaussian
distributions, referred to as ``3D-length'' and ``3D-width'',
respectively. In this method the 3D-width is the most efficient and
robust parameter to discriminate $\gamma$-ray showers from the
hadronic background.

The background subtraction is performed independently in each bin of
the FOV using a maximum likelihood method
(Lemoine-Goumard~et~al.~\cite{HESSWeightingMethod}) based on the
knowledge of the 3D-width distributions of $\gamma$-rays and
background events, respectively. The $\gamma$-ray width distribution
is determined via Monte-Carlo simulations and checked on data from
known point-like sources; the distribution for background events is
obtained from a FOV in the same zenith angle range with no detected
sources. The dependence of these distributions on the bin position in
the FOV is taken into account. The method yields two independent sky
maps, one for $\gamma$-rays and the other for background events. In
contrast to Analysis~1, this method makes no a priori assumption of
azimuthal symmetry in the structure of the background. Furthermore,
the $\gamma$-ray contents of different bins are statistically
independent. This method was applied to the same data sets as for the
analysis presented above; the significance map for $\gamma$-rays,
shown in the lower panel of Figure~\ref{fig:skymap}, is again
compatible with the absence of a signal in the whole FOV.

\subsection{Flux Upper Limits}

\begin{figure}
  \centering
  \includegraphics[width=10.0cm]{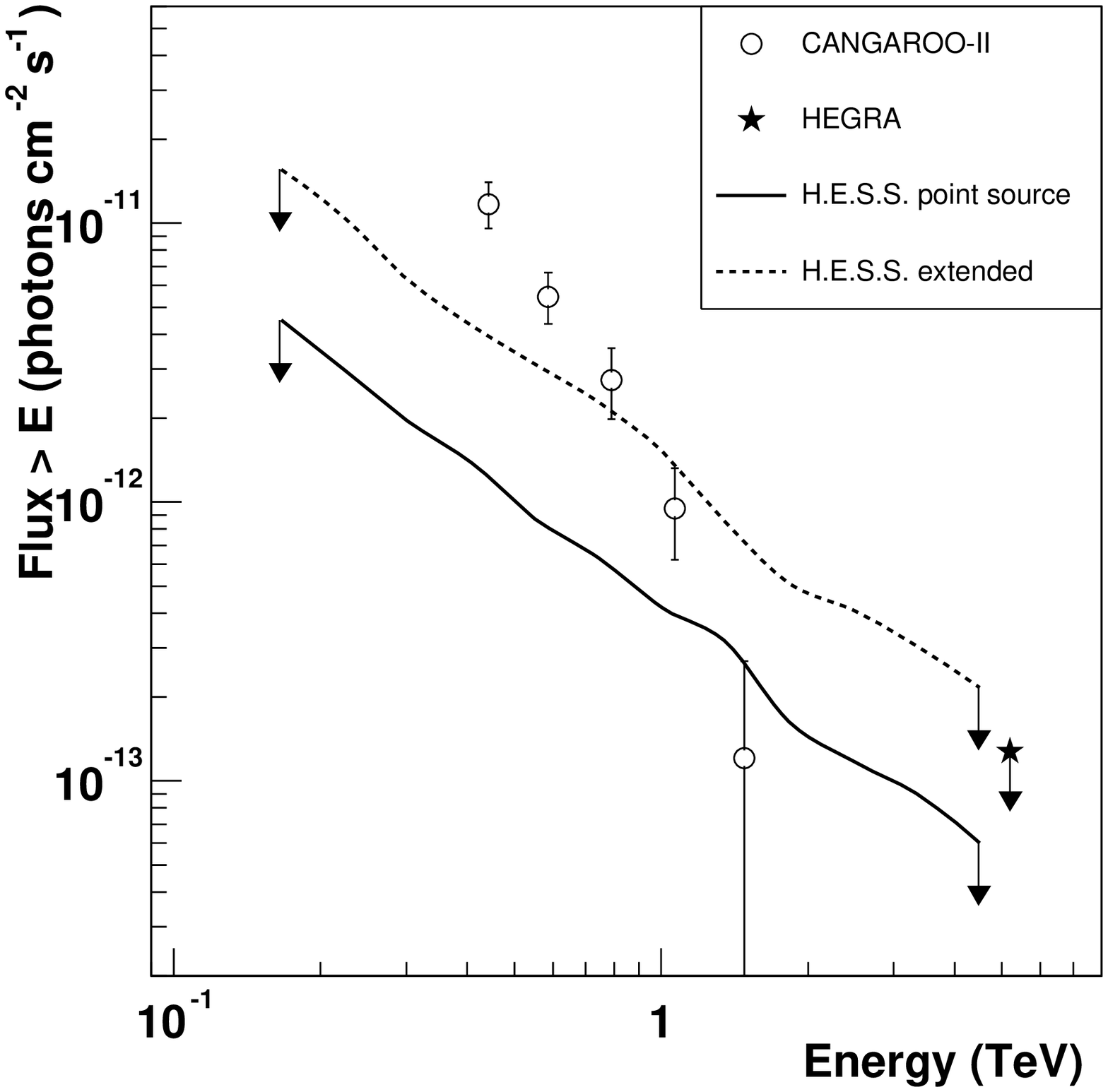}
  \caption{ Upper limits from \hess\ on the integral flux of
    $\gamma$-rays from NGC 253 (99 \% confidence level) from Analysis
    2.  Curves for point like emission (solid lines) and for a source
    of 0.5$^{\circ}$ radius (dashed lines) are shown.  CANGAROO
    integral data points (derived from the differential spectrum given
    by Itoh~et~al.~(\cite{CANGAROOEvidence})) and an upper limit from
    HEGRA~(G\"otting~\cite{HEGRA}) are shown for comparison.  }
  \label{fig:upperlimits}
\end{figure}

Figure~\ref{fig:upperlimits} shows integral flux upper limits
calculated assuming a spectrum of photon index -3.85 (as 
reported by CANGAROO)
and following the method of 
Feldman \& Cousins~(\cite{FeldmanCousins}). Flux limits are shown for 
point-like and extended emission ($0.5^{\circ}$ radius) for 
Analysis 2. Also shown are CANGAROO-II integral data
points derived from the differential spectrum given by
Itoh~et~al. (\cite{CANGAROOEvidence}).
Above 300~GeV we derive 99\% confidence flux upper limits of
$1.9\,\times\,10^{-12}$ photons~cm$^{-2}$ s$^{-1}$ for a point-like source and $6.3\,\times\,10^{-12}$
photons~cm$^{-2}$ s$^{-1}$ for a source of radius 0.5$^{\circ}$. 
The upper limit curves derived from Analysis~1 are rather similar in shape 
and have slightly higher values at 300~GeV: 
$2.2\,\times\,10^{-12}$ photons~cm$^{-2}$ s$^{-1}$ photons~cm$^{-2}$
s$^{-1}$ (point source) and $6.9\,\times\,10^{-12}$ photons~cm$^{-2}$
s$^{-1}$  (0.5$^{\circ}$). 

The non-detection of this object cannot be attributed to a 
mis-pointing of the instrument. The systematic error in the 
pointing error of \hess\ during 2003 was  $<\,1'$~(Gillessen et al.~\cite{Gillessen}). 
Moreover, the correct pointing of the instrument for this particular
dataset has been independently confirmed using the
currents induced by stars in the FOV in the photomultiplier-tube
camera. The limit on the mis-pointing derived from this method is $<\,3'$.

\section{Discussion}

Given the strong signal from NGC\,253 reported by the CANGAROO
collaboration, the non-detection of this object with H.E.S.S. is
surprising.  As the source reported by Itoh~et~al.
(\cite{CANGAROOEvidence}) is extended in nature, time-variability of
the $\gamma$-ray emission cannot be invoked to solve the apparent
contradiction between the \hess\ and the CANGAROO results. In this
situation it is instructive to review the expectations for
$\gamma$-ray emission from this object.

The speculative possibility of detectable $\gamma$-ray emission from
Dark Matter annihilation in the halo of NGC\,253 is discussed in
detail in Enomoto~et~al.~(\cite{Enomoto}). We concur with the view of
these authors that such a scenario cannot produce VHE $\gamma$-rays
with fluxes detectable with current instruments.  The standard picture
of $\gamma$-ray production via CR interactions seems more plausible
given the rapidly growing knowledge about this object in other
wavelength ranges, principly in terms of supernova explosion rates and
the total gas mass in the starburst region.  The relevant CR
interactions are: i) inelastic collisions of relativistic nuclei with
thermal gas nuclei producing $\pi^{0}$ mesons which decay into
$\gamma$-rays ii) Inverse Compton (IC) collisions of relativistic
electrons with ambient low-energy photons iii) Bremstrahlung
collisions of such electrons with gas nuclei.  The $\gamma$-ray flux
from these processes is proportional to the density of these
relativistic particles (CRs). The CR density is determined by the
strength of their accelerators in the SB region, their propagation
modes, and their energy losses.  In particular, propagation of CRs via
convection in the SB wind is a key process that is usually not
considered for our own galaxy.  In addition, the nature of particle
diffusion in the SB is likely modified by non-linear effects induced
by the CRs themselves.

A self-consistant picture for the $\gamma$-ray production 
is discussed below. It turns out that the strong localisation of CR production 
and the high gas mass of the SB region lead to a rather 
simple prediction for the VHE $\gamma$-ray flux, 
independant of the further details of the SB configuration.

\subsection{Starburst parameters for NGC\,253}

The enhanced CR density and the very high gas and photon
densities in the starburst (SB) regions are related to each other: the
high gas densities are necessary for the high star formation rate which
produces massive stars of high photon luminosity. Their subsequent supernova
(SN) explosions heat the gas and generate CRs. Together the heating of the
gas and the high CR density drive a mass outflow in the form of a
galactic wind.  

NGC\,253 has a central SB region which is very bright in radio, far
infrared, and X-ray frequencies. Following
Weaver~et~al.~(\cite{Weaver}) we assume it to be a cylinder of radius
$R\,=\,150$~pc and height $H\,=\,60$~pc in the plane of the galaxy,
with an outflow that is collimated by a massive molecular torus into
the vertical direction, perpendicular to the disk.  The outflow
velocity $V_{\mathrm{wind}}$ is not very well known, with ``{\it
minimum} reasonable values between 300 and 600~km\,s$^{-1}$''
(Strickland~et~al.~\cite{Strick2}). However, values for the outflowing
SN ejecta as high as 3000~km\,s$^{-1}$ are considered possible.  For
the purposes of our estimates we assume here $V_{\mathrm{wind}}=
500$~km\,s$^{-1}$.

The volume of the SB region is then $V_{\mathrm{SB}} \approx 4.2\times
10^{6}$~pc$^3$.  The SN rate in this small nuclear volume is estimated
as $\nu_{\mathrm{SN}}= 0.03$~yr$^{-1}$, about as high as that of our
entire Galaxy (cf. Engelbracht~et~al.~\cite{Engelbracht}), but values as
high as 0.08~yr$^{-1}$ have also been estimated~(van~Buren \&
Greenhouse~\cite{vanBuren}; Paglione~et~al.~\cite{Paglione}). The rest
of the galaxy is thought to have a SN rate of the same magnitude,
distributed over the disk. The total (dominantly molecular) mass
$M$ in the nuclear region is rather uncertain: $3\times
10^7 \leq M/M_{\odot} \leq 1.4 \times 10^8$~
(Mauersberger~et~al.~\cite{Mauersberger},
Engelbracht~et~al.~\cite{Engelbracht}). We assume $M =
6\times 10^7~M_{\odot}$ which corresponds to the preferred value of
Engelbracht~et~al.~(\cite{Engelbracht}) and gives a density of 
580~protons~cm$^{-3}$ in the nuclear region.

\subsection{Transport of VHE particles}

The non-thermal properties of such an SB system are quite complex, as
the very extended radio synchrotron emission shows. This radio
emission is intimately connected to the large expanding halo in
thermal X-rays and optical $\mathrm{H}_{\alpha}$-line
photons. Nevertheless, the TeV $\gamma$-ray emission can be discussed
in a fairly straightforward manner. For this purpose,
let us assume that particle diffusion can be calculated from the theory of 
the wind from our Galaxy (Ptuskin~et~al.~\cite{Ptuskin}), scaled to the SB 
parameters of NGC\,253.
We point out that in the theory of Ptuskin~et~al. the
diffusion coefficient in the Galactic wind is assumed to be due to
Alfv\'en waves, excited by the particles themselves, as they stream away
from their sources in the disk. The wave amplitudes are limited by
nonlinear wave damping on the thermal wind plasma.

To estimate the spatial extent and magnitude of the $\gamma$-ray
emission as a result of the SB, we first consider the time scales for
energetic particle transport and their collisional/radiative energy
losses. For protons the convection time $t_{conv}=
H/V_{\mathrm{wind}}\approx 1.2\times 10^5$~yr, whereas the loss time
in the nuclear region due to inelastic collisions with nuclei of the
gas amounts to $t_{\mathrm{loss}} = 8.6\times 10^{4}$~yr.

The diffusion vertically to the disk is given by the diffusion
coefficient $\kappa =
(B/B_{\mathrm{gal}})(\epsilon/\epsilon_{\mathrm{gal}})^{-1}
\kappa_{\mathrm{gal}}(E)$, where above a few GeV
$\kappa_{\mathrm{gal}}(E) \approx 9 \times
10^{29}~(E/1\mathrm{TeV})^{1.1}~\mathrm{cm}^{2} \mathrm{s}^{-1}$ is
consistent with the mean matter thickness traversed by the CRs in our
Galaxy (and its rigidity dependance), before they are ultimately
convected away into extragalactic space
(Ptuskin~et~al.~\cite{Ptuskin})
\footnote{In our galaxy $V_\mathrm{wind}\propto s$ for $s<$15~kpc
where $s$ is the height above the disk, The distance $s_{\star}$
within which the energetic particles can diffusively return to the
disk to produce secondaries is given by
$\kappa_{\mathrm{gal}}(E)/s_{\star}^{2}=V_{\mathrm{wind}}(s_{\star})/s_{\star}$
(the diffusion-convection boundary). Thus
$s_{\star}\propto\kappa_{\mathrm{gal}}^{1/2}(E)\propto\,E^{0.55}$ for
our galaxy. The probability for a particle to cross the disk, and thus
the mean matter thickness traversed by particles of energy $E$ is
inversely proportional to $s_{\star}$.}.
%In this scaling relation $\epsilon$ denotes the spatial energy flux
%density of CRs released into the SB region. 
In this scaling relation $\epsilon \propto
\nu_{\mathrm{SN}}/$\emph{(CR source surface area)} denotes the spatial
energy flux density of CRs released into the SB region. In detail,
$\epsilon = 4 \pi {\mathrm c} \int_{0}^{\infty} p^3 Q(p) dp$, where
$Q(p)$ is the CR source power density, as a function of particle
momentum $p$, per unit surface area of the source region.  As we shall
see, in this case $t_{\mathrm{loss}}$ is smaller than both
$t_{\mathrm{conv}}$ and $t_{\mathrm{diff}}$ (which is defined
below). This results in an {\it upper limit} for the $\gamma$-ray
emission from NGC\,253 for a given energy input into CRs.

To determine $\epsilon$ for NGC\,253 we note that the cylindrical SB
region has a radius of 150~pc, according to the above model, whereas
the corresponding region of our Galaxy, the disk, has a radius of
15~kpc. Therefore, assuming $\nu_{\mathrm{SN}}= 0.03$~yr$^{-1}$ in the
SB as in the Milky Way, yields $\epsilon/\epsilon_{\mathrm{gal}}=
10^{4}$.  For the magnetic field $B$ in the SB region we chose $B\leq
270\,\mu$G, where $270\,\mu$G corresponds to the equipartition value
derived by Weaver~et~al.~(\cite{Weaver}) which we consider the maximum
possible field in the SB region. With a Galactic field strength
$B_{\mathrm{gal}}=3\,\mu$G this massive CR production reduces the
diffusion coefficient to $\kappa\,\leq 9\times
10^{-3}\,\kappa_{\mathrm{gal}}$. The corresponding diffusion time
$t_{\mathrm{diff}} = H^2/\kappa \geq 1.3\times 10^5$~yr for a 1~TeV
particle also exceeds $t_{\mathrm{loss}}$. This means that CR protons
with energies up to about 1~TeV behave {\it calorimetrically}:
i.e. they lose all their energy by inelastic collisions before being
able to leave the SB region. However, the increase of $\kappa$ with
particle energy makes the particles with energies above several TeV
{\it diffusion dominated}, which implies that they escape the SB
region before interacting. The turnover point is at $\gamma$-ray
energies of a few hundred GeV.

For TeV electrons the loss time is estimated as follows: the enormous
far infrared (FIR) SB-luminosity $L_{\mathrm{SB}}=1.1\times 10^{10}
L_{\odot}$ deduced by Engelbracht~et~al.~(\cite{Engelbracht}) implies
a radiation energy density $U_{\mathrm{rad}}^{\mathrm{SB}} \approx
500$~eV\,cm$^{-3}$.  This leads to an inverse Compton (IC) loss time
$t_{\mathrm{loss}}^{\mathrm{IC}} \approx 180\,
(E/4\mathrm{TeV})^{-1}$~yr for the electrons of energy $E\approx
4$~TeV that produce 1 TeV $\gamma$-rays. The IC scattering is here
considered in the Thomson limit which applies for FIR target photons.
Therefore, comparing with the proton emission, the SB region acts even
more calorimeterically for the emission of TeV $\gamma$-rays via IC
scattering.

\subsection{VHE $\gamma$-ray emission}

The hadronic $\gamma$-ray emission in this scenario simply corresponds
to the total amount of CR energy created.  The $\gamma$-ray energy
spectrum corresponds to that of the SNR sources, assumed here to be
$dN_{\mathrm{CR}}/dE\,\propto\,E^{-2.1}$.  In this picture diffusion
is irrelevant below a few TeV, and the total CR energy
$E_{\mathrm{CR}}$ in the SB region amounts to
\begin{eqnarray}
E_{\mathrm{CR}} \approx 1.5\times 10^{53}~\mathrm{erg} 
\left[\frac{\nu_{\mathrm{SN}}}{0.03 \mathrm{yr}^{-1}} 
\,\frac{\Theta E_{\mathrm{SN}}}{10^{50}\mathrm{erg}}\,\frac{t_{\mathrm{eff}}}{5\times 10^{4} 
\mathrm{yr}}\right],
\end{eqnarray}
where
$t_{\mathrm{eff}}^{-1}=t_{\mathrm{loss}}^{-1}+t_{\mathrm{conv}}^{-1}+t_{\mathrm{diff}}^{-1}$,
and $\Theta<1$ is the CR energy production efficiency for an average
SN event with a mechanical energy release $E_{\mathrm{SN}}$. The CR
energy density $E_{\mathrm{c}}^{253}=E_{\mathrm{CR}}/\mathrm{Volume}$
amounts to $E_{\mathrm{c}}^{253}\approx 650$~eV\,cm$^{-3}$. This has
to be compared to the Galactic value
$E_{\mathrm{c}}^{\mathrm{gal}}\sim 1$~eV\,cm$^{-3}$.  Assuming the
radiating TeV particles to freely penetrate the gas in and around
their sources -- a premise which would become more and more
questionable at lower energies on account of the shorter scattering
mean free path $\lambda=\,3\kappa(E)/c$ -- the energy spectrum of the
hadronic $\gamma$-ray flux can be calculated from the following
relation (cf. V\"olk~et~al.~\cite{Voelk2}):
\begin{eqnarray}
F_{\gamma}(>E) = 1.7\times 10^{-13}\,\mathrm{photons}\,\,\mathrm{cm}^{-2}\,\mathrm{s}^{-1}\,\times\hspace{16mm}\nonumber\\
\hspace{9mm}\left[\left(\frac{E}{1 \mathrm{TeV}}\right)^{-1.1}
\left(\frac{E_{\mathrm{c}}}{1 \mathrm{eV}\,\mathrm{cm}^{-3}}\right)
\left(\frac{d}{1 \mathrm{Mpc}}\right)^{-2} \left(\frac{M}{10^{9}M_{\odot}}\right)\right].
\end{eqnarray}

Substituting our preferred values for $d$, $M$, and $E_{\mathrm{c}}$,
as calculated above, we obtain $F_{\gamma}(>E) = 9.8\times
10^{-13}\,\times\,\delta\,\times\,(E/1
\mathrm{TeV})^{-1.1}$~photons~cm$^{-2}$ s$^{-1}$ up to a few hundred
GeV, and significantly falling off at higher $\gamma$-ray energies.
The factor $\delta$ denotes the uncertainty of the assumed parameters,
mainly $\nu_{\mathrm{SN}}$, $M$, and $\Theta$. Assuming that $\Theta
\approx 0.1$ and the average $E_{\mathrm{SN}}$ = $10^{51}$~erg, we
obtain $0.5<\delta<5$.

As long as $M\geq 6\times 10^7 M_{\odot}$ we have
$t_{\mathrm{eff}}\approx t_{\mathrm{loss}}\propto V_{\mathrm{SB}}/M$.
Since $E_{\mathrm{c}}\propto\,t_{\mathrm{loss}}/V_{\mathrm{SB}}$,
$F_{\gamma}$ (where $F_{\gamma}\propto\,E_{\mathrm{c}}\cdot\,M$) then
becomes independent of $M$, $V_{\mathrm{SB}}$, $t_{\mathrm{conv}}$ and
$t_{\mathrm{diff}}$, and is only proportional to the energy input per
SN explosion $\Theta E_{\mathrm{SN}}$ and the SN rate
$\nu_{\mathrm{SN}}$, apart from the geometrical factor $d^{-2}$. This
is the consequence of the calorimetric behaviour of the system.

The CR electron energy produced in the SNR sources of the SB can be
assumed to amount to no more than a fraction $\sim 10^{-2}$ of the
energy generated in nuclear CR particles, as is the case in our
galaxy. The contribution of IC scattering to the $\gamma$-ray emission
in the VHE region can therefore be safely disregarded.  The above
hadronic flux value, with $\delta$=1, is similar to -- in fact
somewhat larger than -- the H.E.S.S. upper limit for the VHE
$\gamma$-ray flux. At the most relevant energy of about 200~GeV the
predicted flux is about a factor of 3 larger than the H.E.S.S.  limit
for an assumed point source. This demonstrates the physical relevance
of the H.E.S.S. upper limit measurement.

Our discussion rests on the assumption that the propagation properties
of energetic particles in our own Galaxy and its halo can be scaled to
the SB parameters of NGC\,253. If this assumption does not hold, then
some fraction of the accelerated particles will escape the galaxy
before radiating, and the predicted $\gamma$-ray flux from NGC\,253
will decrease relative to the estimate given here.

The emission from the rest of the galaxy may be spatially extended.
However, assuming a situation roughly equal to that of the Milky Way,
where the amount of matter traversed by a TeV particle corresponds to
less than 1.5\% of the hadronic interaction length, the majority of
the protons will escape and will not transfer their energy to
$\gamma$-rays.  Therefore their total $\gamma$-ray emission is
substantially smaller than that from the SB region. The IC
$\gamma$-ray flux from the associated electrons is still negligible
compared to the CANGAROO flux.

This scenario leads to the prediction that VHE $\gamma$-ray emission
from the SB in NGC\,253 will be highly localised, with an extent of
less than 30 arcseconds. Such emission would appear point-like for
current VHE $\gamma$-ray instruments. As discussed earlier, extended
emission from the galaxy as a whole is expected to be rather faint in
comparison to the flux reported by the CANGAROO collaboration.  In
addition the expected spectrum should be rather hard as discussed
earlier, again in contrast to the reported CANGAROO spectrum.

\section{Conclusions}

The recently completed H.E.S.S. instrument is currently the most
sensitive detector of $\gamma$-rays at VHE energies. The non-detection
of NGC\,253 by H.E.S.S. is surprising given the rather high flux
previously claimed by~Itoh~et~al.~(\cite{CANGAROODetection}). From
theoretical considerations of $\gamma$-ray production in starburst
galaxies it appears that the H.E.S.S. limit is close to the expected
flux from the NGC\,253 starburst, already excluding extreme parameter
values ($\delta>1$). However, a diffusive transport away from the
starburst region that is faster than assumed here cannot be excluded
experimentally. The possible $\gamma$-ray flux would then be
correspondingly reduced.

It also seems likely that VHE $\gamma$-ray emission from NGC\,253 should
be rather localised (indeed point-like for current detectors), 
given the characteristics of the starburst measured at other wavelengths.
A hard $\gamma$-ray photon index ($\approx 2$) would also be expected.

Despite this non-detection, NGC\,253 remains one of the most promising
starburst galaxies for detection in very high energy
$\gamma$-rays. The likely calorimetric character of this system makes
deeper observations an interesting proposition.

%_________________________________________________________________
\section*{Acknowledgements}

The support of the Namibian authorities and of the University of Namibia
in facilitating the construction and operation of H.E.S.S. is gratefully
acknowledged, as is the support by the German Ministry for Education and
Research (BMBF), the Max Planck Society, the French Ministry for Research,
the CNRS-IN2P3 and the Astroparticle Interdisciplinary Programme of the
CNRS, the U.K. Particle Physics and Astronomy Research Council (PPARC),
the IPNP of the Charles University, the South African Department of
Science and Technology and National Research Foundation, and by the
University of Namibia. We appreciate the excellent work of the technical
support staff in Berlin, Durham, Hamburg, Heidelberg, Palaiseau, Paris,
Saclay, and in Namibia in the construction and operation of the equipment.
We thank V.N. Zirakashvili for discussions regarding the
transport properties of starburst galaxies.

%_________________________________________________________________

\end{document}